\documentclass[graybox,vecphys]{svmult}

\usepackage{type1cm}        
\usepackage{makeidx}         
\usepackage{graphicx}        
\usepackage{multicol}        
\usepackage[bottom]{footmisc}

\usepackage{newtxtext}       %
\usepackage{newtxmath}       
\usepackage{url}

\makeindex             

\usepackage{units} 

\providecommand{\bc}{\begin{center}}
\providecommand{\ec}{\end{center}}
\providecommand{\be}{\begin{equation}}
\providecommand{\ee}{\end{equation}}
\providecommand{\bea}{\begin{eqnarray}}
\providecommand{\eea}{\end{eqnarray}}
\providecommand{\bdm}{\begin{displaymath}}
\providecommand{\edm}{\end{displaymath}}
\providecommand{\bdma}{\begin{eqnarray*}}
\providecommand{\edma}{\end{eqnarray*}}
\providecommand{\ba}{\begin{eqnarray*}}
\providecommand{\ea}{\end{eqnarray*}}
\providecommand{\bi}{\begin{itemize}}
\providecommand{\ei}{\end{itemize}}
\providecommand{\benum}{\begin{enumerate}}
\providecommand{\eenum}{\end{enumerate}}

\providecommand{\twoCases}[4]{
  \left\{ 
    \begin{array}{ll} 
      #1 & #2 \\
      #3 & #4 
    \end{array} 
  \right.
}

\providecommand{\text}[1]{{\mbox{ #1}}}

\usepackage{tikz}

\providecommand{\fig}[2]{
   \begin{center}
     \includegraphics[width=#1]{#2}
   \end{center}
}

\providecommand{\ablpart}[2]{\frac{\partial #1}{\partial #2}}  
\providecommand{\abl}[2]{\frac{{\rm d} #1}{{\rm d} #2}}  

\providecommand{\sub}[1]{_{\rm #1}}
\renewcommand{\sup}[1]{^{\rm #1}}

\begin{document}

\title*{The intelligent agent model -- a fully two-dimensional
  microscopic traffic flow model}
\titlerunning{The intelligent agent model}

\author{Martin Treiber and Ankit Anil Chaudhari}
\authorrunning{Martin Treiber and Ankit Anil Chaudhari}

\institute{Martin Treiber \at TU Dresden, \email{martin.treiber@tu-dresden.de}
\and Ankit Anil Chaudhari \at TU Dresden, \email{ankit_anil.chaudhari@tu-dresden.de}}
%
%
\maketitle

\vspace{-5em}

\abstract{Recently, a fully two-dimensional microscopic traffic flow model for
lane-free vehicular traffic flow has been proposed [Physica A, 509,
pp. 1-11 (2018)]. In this contribution, we generalize this model to
describe any kind of human-driven directed flow including
lane-based vehicular flow, lane-free mixed traffic, bicycle traffic,
and pedestrian flow. The proposed
intelligent-agent model (IAM) has the same philosophy as the
well-known social-force model (SFM) for pedestrians but the interaction
and boundary forces are based on car-following 
models making this model suitable for higher speeds. Depending on the underlying car-following model, the IAM
includes anticipation, response to relative velocities, and
accident-free driving. 
When adding a suitable floor field, the
IAM reverts to an integrated car-following and lane-changing model
with continuous lane changes.
We simulate this model
in several lane-based and lane-free environments in various
geometries with and without obstacles. We observe that the model
 produces accident-free traffic flow 
reproducing the observed self-organisation phenomena.
}

\section{Introduction}
\label{sec:1}
Microscopic traffic flow models traditionally describe the flow in a
continuous longitudinal and discrete lateral dimension in form of
car-following and lane-changing models
\cite{TreiberKesting-Book}.
In traffic flow simulators such as VISSIM~\cite{VISSIM} or
SUMO~\cite{SUMO2011}, both components are integrated into a common
model. These traditional models, however, are not able to describe
disordered lane-free traffic flow consisting of a wide variety of
vehicle sizes and properties which is typically observed in developing
countries. This type of flow is characterized by a continuous lateral
degree of freedom requiring a fully two-dimensional model. Generalizing
lane-changing models with incentives in
terms of acceleration differences~\cite{Hidas-02,MOBIL-TRR07}, a first
fully two-dimensional model for mixed vehicular flow with lateral forces
proportional to longitudinal acceleration shears has been
proposed, the mixed-traffic model (MTM)~\cite{Kanagaraj2018self}.

However, this model only reacts to vehicles in the front and also has
a rather complicated formulation of the lateral forces making it hard
to calibrate. In contrast, the social-force model (SFM) for
pedestrians by Helbing and  Moln{\'a}r~\cite{SFM} has a clean
formulation. There, the acceleration of each pedestrian is given by
the superposition of the free-flow social force, the interaction
forces exerted by the pedestrians nearby, and the repulsive forces of the
boundaries. However, the SFM cannot be applied to vehicles, cyclists
or other self-driven agents with a higher speed since it does not
contain the kinematic
constraints of a limited acceleration, is not crash free, and does not revert to a
plausible car-following model in case of single-file traffic. This is
also the case for other pedestrian flow models~\cite{chen2018social}.

On the other hand, there exist several car-following models for mixed
traffic which, however, only modify the model parameters depending on
the lateral offset or the kind of leaders and
followers~\cite{asaithambi2018study,madhu2023vehicle} while not
incorporating any lateral dynamics itself.

In this contribution, we propose the intelligent-agent model (IAM)
which is an integration of the (simplified) MTM for vehicular traffic
and the SFM for
pedestrians. As the MTM, it is based on a car-following model to which
it reverts for single-file traffic and from which it inherits all the
properties for a safe high-speed motion. It also contains aspects of
the social-force model such as the superposition of the social forces
of all active agents nearby (including the back) with a directional
weighting. 
Optionally, we also introduce floor fields to transform the originally
lane-free formulation to a lane-based environment. In this case, the
IAM reverts to an integrated car-following and lane-changing model
with continuous lane changes.

In Sect.~\ref{sec:model}, we formulate our proposed
model IAM. In Section~\ref{sec:sim}, we simulate it with the
underlying Intelligent-Driver-Model~\cite{Opus,traffic-simulation} 
in several lane-based and
lane-free environments with mixed vehicular-bicycle traffic and
pure bicycle traffic. We conclude with a discussion in
Sect.~\ref{sec:concl}.

\section{Model specification}
\label{sec:model}
As the SFM, the proposed IAM is based on social forces which can be
subdivided into the self-driving force to reach the destination with a
desired speed, the
interaction forces with other moving and standing objects,
and boundary forces to keep the agents in the driveable or walkeable
area,
\be
\label{IAM}
\abl{\vec{v}_i}{t}=\vec{f}\sup{self}_i(\vec{v}_i)
 +\sum_j \vec{f}\sup{int}_{ij}(\vec{r}_i,\vec{v}_i, \vec{r}_j,\vec{v}_j)
 +\sum_b \vec{f}_{ib}.
\ee
 Unlike the SFM, the IAM is dedicated to directed flows where a
local axis of the road or pathway can be defined. Furthermore, the IAM
is anisotropic with a distinctively different dynamics for 
 the longitudinal direction (parallel to
the axis) and the lateral one. We denote the components of the
longitudinal and lateral position vector with $\vec{x}=(x,y)'$, and
decompose the velocity and force vectors by $\vec{v}=(v,w)'$ and
$\vec{f}=(f,g)'$, respectively. Setting the agent's mass $m=1$, the
forces also denote the accelerations, $\vec{f}=(\dot{v},\dot{w})'$.

\subsection{Self-driving forces}
\label{sec:self}
The longitudinal part of the self-driving force is derived from the
underlying car-following (CF) which we characterize by the general
acceleration function 
$\dot{v}\equiv \abl{v}{t}=f\sup{CF}(s,v,v_l)$ depending on the
(bumper-to-bumper) gap $s$ to the
leader, the subject's speed $v$, and the leading speed $v_l$.
We derive the self-driving force by decomposing this model as 
\be
\label{f-free}
f\sup{CF}(s,v,v_l)=f\sup{self}(v)+f\sup{CF,int}(s,v,v_l),
\quad f\sup{self}(v)=f\sup{CF}(s\to\infty,v,v),
\ee
where 
$f\sup{CF,int}=f\sup{CF}-f\sup{self}$ 
denotes the CF force in case of a strictly single-file traffic. 
Notice that the desired speed $v_0$ is given in terms of the implicit relation
$f\sub{CF}(s\to\infty, v_0,v_0)=0$.

Since we assume situations where the lateral velocity component is
of the order of the pedestrian speed, the free-flow velocity
relaxation term of the SFM is 
appropriate resulting in the lateral self-driving force
\be
\label{g-free}
g\sup{self}(w)=\frac{w_0-w}{\tau_y},
\ee
where the lateral desired speed $w_0=v_0e_y\sup{dest}$ is nonzero if the
unit vector
$\vec{e}\sup{dest}$ to the destination is not parallel to the road
axis, for example, when entering or leaving the road or
bikeway.

\subsection{Interaction forces}
According to Eq~\eqref{IAM}, the interacting force is a
superposition of the forces $\vec{f}_{ij}\sup{int}=
\left(f_{ij}\sup{int}, g_{ij}\sup{int}\right)'$ of the nearby agents
including standing objects, e.g., the stopping line of a
red traffic light which is represented by a very wide and very short
virtual vehicle with the outline of the stopping line. In the
following, we define the longitudinal and lateral forces exerted by an
agent/object $j$ as a function of $\vec{x}_j$ and $\vec{v}_j$.

\subsubsection{Longitudinal interaction forces}
We assume that the longitudinal interaction is that of the CF model
whenever the follower's occupancy area laterally encroaches that of the
leader. If the lateral speeds are negligible with respect to the 
longitudinal ones, encroachment is given if the absolute value of the lateral
offset $\Delta y=y_j-y_i$ is less than the average agent/object with
$\overline{W}=(W_i+W_j)/2$. If no encroachment is given, i.e., the lateral
gap $s_y=|\Delta y|-\overline{W}>0$, the longitudinal
force decreases exponentially with $s_y$ as in the SFM. This results
in
\be
\label{f-int}
f_{ij}\sup{int}(\vec{x}_i, \vec{x}_j, \vec{v}_i, \vec{v}_j)
=\alpha(s_y)f\sup{CF}(s_x,v_i,v_j)
\ee
where
\be
\label{f-int-defs}
s_x=x_j-x_i-L_j, \quad s_y=|y_j-y_i|-\overline{W}, \quad
\alpha(s_y)=\min \left(1, e^{-|s_y|/s_{0y}}\right).
\ee
The underlying CF model must be specified in a way that negative gaps $s_x$
(i.e., collisions) result in a maximum deceleration
(e.g. $\unit[9]{m/s^2}$). Furthermore, we assume a longitudinal force
of zero if $s_x<0$ but $s_y>0$, i.e., the agent $j$ drives in
parallel.

\runinhead{Interactions from followers} 
As in the SFM, we also include ``pushing'' interactions from the
followers ($x_j<x_i$) which are weakend by
the SFM anisotropy factor $\lambda\le 1$,
\be
\label{f-int-push}
f_{ij}\sup{push}=-\lambda f_{ji}\sup{int}
\ee
With $\lambda=1$, we would have a momentum conserving dynamics
(``actio=reactio'') but no longer a defined fundamental diagram, i.e.,
a steady-state macroscopic flow-density relation. Plausible values (of
the order of $\lambda=0.1$) increase the flow efficiency without leading
to more critical situations.

\runinhead{Total longitudinal interaction}
Reacting to more than one leader or follower can lead to inconsistent
results if there are large size differences between the agents. For
example, three cyclists driving in parallel and followed by a truck
will not exert on the truck driver the triple social force. Rather,
the truck driver reacts to the slowest and/or nearest bicycle,
only. Therefore, the total longitudinal interaction force is given by
\be
\label{f-int-tot}
f_i\sup{int}=\min_{j, x_j\ge x_i}f_{ij}\sup{int}
            +\max_{j, x_j<x_i}f_{ij}\sup{push}.
\ee
Notice that, even in single-file traffic, the selected leader needs
not to be the 
immediate leader but can also be a red traffic light further downstream.

\subsubsection{Lateral interaction forces}
Generalizing the incentive criterion of the lane-changing model MOBIL,
the lateral incentive exerted by agent $j$ is proportional to the
shear of the longitudinal force from this agent. Formulating the
incentive in terms 
of an induced desired lateral velocity $w_{0i}$, we have
\be
\label{w0int}
w_{0,ij} \propto \partial f_{ij}\sup{int}/\partial y_i
\ee
leading via~\eqref{f-int}, as in the SFM, to an exponential decrease
of the repulsion  if there is no lateral overlap. However, if there is an
overlap, the longitudinal force does not change with $y_i$ and the
lateral incentive is zero. To break this tie, we assume that the lateral
incentive linearly increases with the lateral offset in the
overlapping region. Inserting~\eqref{w0int} into~\eqref{f-int} and describing the lateral dynamics, as in
the SFM, as a relaxation process, we finally obtain
\be
\label{g-int}
g\sup{int}_{ij}=\frac{w_{0,ij}-w_i}{\tau}, \quad
w_{0,ij}=\sigma f\sub{CF}\sup{int}
\twoCases{\Delta y/\overline{W}}{\ \ |\Delta y| \le \overline{W},}
{\text{sign}(\Delta y) e^{-{s_y/s_{0y}}}}{\ \ \text{otherwise}.}
\ee
Unlike the longitudinal case, we assume no shielding, so the total
lateral interaction force is given by the sum of the forces exerted by
all surrounding agents with that of the followers reduced by the 
anisotropy factor $\lambda$.


\begin{table}[!t]
\caption{Model parameters of the IAM in addition of the parameters of
  the underlying CF model}
\label{tab:param}       
%
%
\begin{tabular}{p{6cm}p{2cm}p{3cm}}
\hline\noalign{\smallskip}
Parameter & Meaning & Typical value  \\
\noalign{\smallskip}\svhline\noalign{\smallskip}
SFM transversal relaxation time & $\tau_y$ & \unit[1]{s}\\
Attenuation width & $s_{0y}$ & \unit[0.3]{m}\\
Boundary attenuation width & $s_{B0}$ & \unit[0.2]{m}\\
SFM anisotropy parameter & $\lambda$ & 0 - 0.2 \\
Lateral sensitivity & $\sigma$ & \unit[1]{s} \\
long. deceleration at the boundary & $\hat{f}\sub{B}$ & $\unit[0.2]{m/s^2}$ \\
lat. acceleration at the boundary & $\hat{g}\sub{B}$ & $\unit[5]{m/s^2}$ \\
\noalign{\smallskip}\hline\noalign{\smallskip}
\end{tabular}
\end{table}
%

\begin{figure}[b]
\sidecaption[t]
\includegraphics[width=7.5cm]{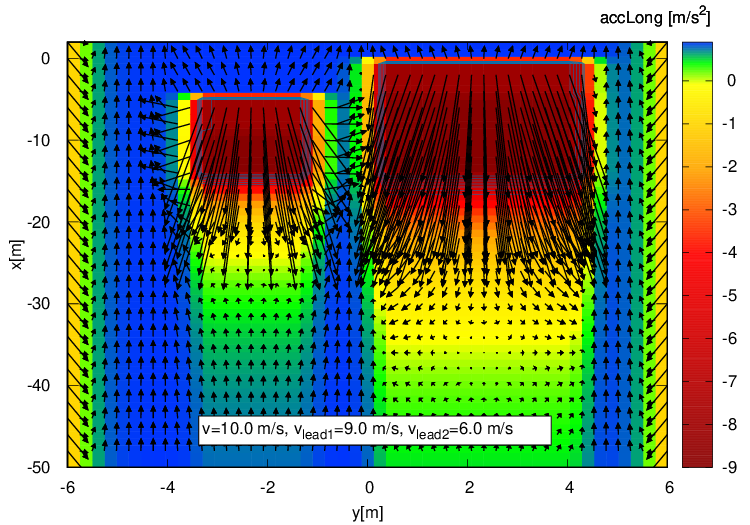}
\caption{\label{fig:accField}Force (acceleration) vector field of all
  forces of 
  Section~\ref{sec:model} for two leaders and road boundary for
  arbitrary positions ($x,y$) of the follower. Notice that leader~2
  (to the right) is bigger, further away, and slower}
\label{fig:accvector} 
\end{figure}

\subsection{Boundary forces and floor fields}

In principle, a boundary of a driveable area can be modelled by a
series of long obstacles. However, an equivalent dedicated approach is
more efficient. When driving near the road boundary or even
transgressing it partly, there is not only a strong social force
towards the road center but also a weak decelerating force in the
longitudinal direction:
\be
\label{f-boundary}
\vec{f}\sub{B}(s_{y})=\alpha\sub{B}(s_{y})
 \left(-\hat{f}\sub{B} \frac{v}{v_0}, \
 \pm \hat{g}\sub{B} \right)', \
\alpha\sub{B}(s)=\min\big(1, e^{-s/s\sub{B0}}\big),
\ee
where $s_{y}$ is the gap between the boundary and the vehicle
($s_{y}<0$ if the boundary is transgressed) and
$\hat{f}\sub{B}$, $\hat{g}\sub{B}$ and $s\sub{B0}$ are model parameters (cf
Table~\ref{tab:param}). 



Finally, to obtain the resulting acceleration vector caused by the
boundaries, all forces from the left and right
boundary are added up (Fig.~\ref{fig:accField}).

\runinhead{Floor fields} For lane-based flow, we add a periodic floor
field:
\be
\label{floorfield}
g\sub{lane}(x,y)=-\ablpart{\Phi}{y}, \quad
\Phi(x,y)=\pm \Phi_0 \cos(2\pi y/W\sub{lane})
\ee
where plus applies for an
even lane number (a lane separating line is on the directional road
axis) and  minus for an uneven number. The maximum induced lateral
acceleration is given by $2\pi\Phi_0/W\sub{lane}$.

\section{Simulation and validation of lane-based and lane-free scenarios}

\label{sec:sim}

In this section, we simulate the IAM for two different scenarios: Lane
based vehicular traffic (Sect.~\ref{sec:Athens}) and lane-free bicycle
traffic (Sect.~\ref{sec:bike}). Both simulations are validated by
empirical data.

\begin{figure}[b]
\fig{0.95\textwidth}{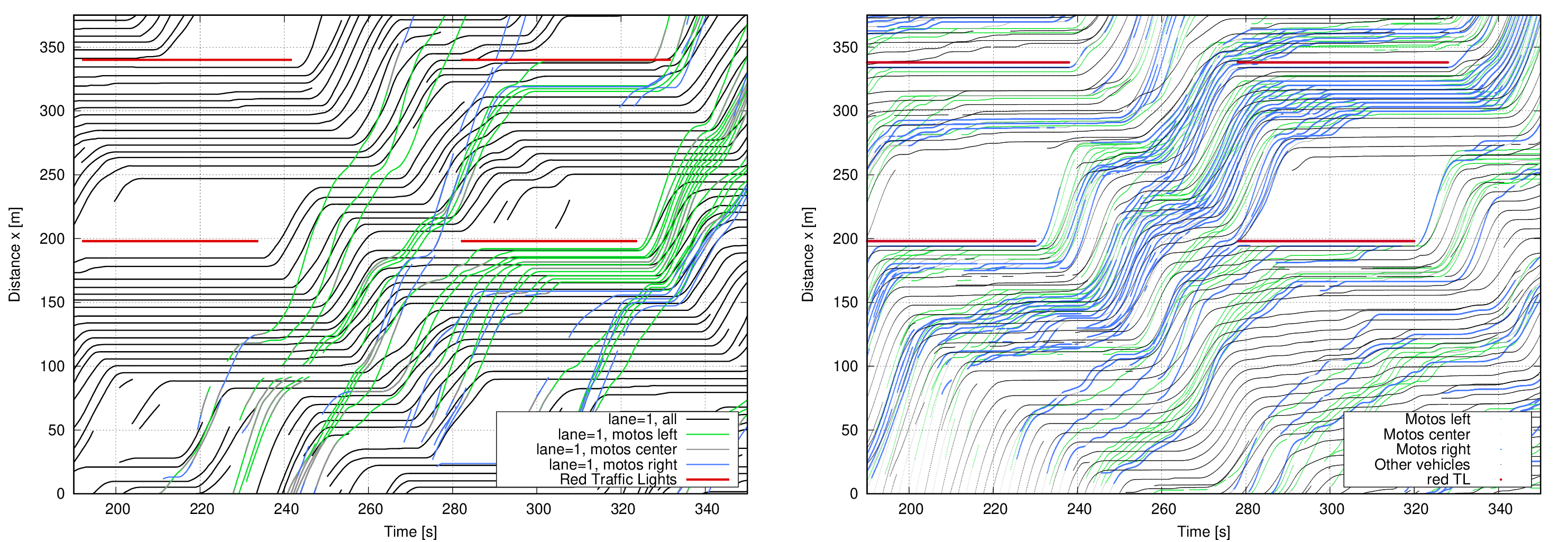} 
\caption{\label{fig:Athens-xt}Left: Trajectory data  taken from the
  open-data project pNEUMA~\cite{pNEUMA}. Right: simulation of the
  IAM with floor fields. In both images, the motorcycles are
  color-coded whether they drive in the lane (gray) or between the lanes to
  the left or right (green and blue, respectively).}
\end{figure}

\begin{figure}[b]
\fig{0.95\textwidth}{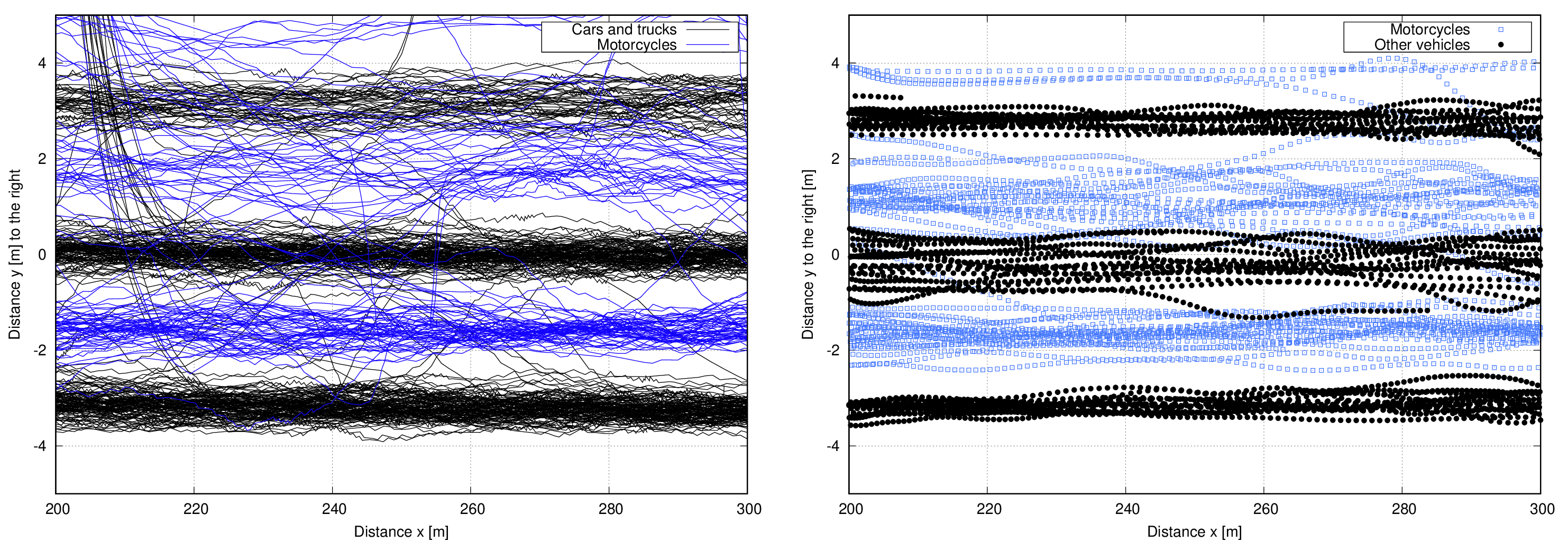}
\caption{\label{fig:Athens-xy}Cross section on the $x$-$y$ plane of
  the same observed and simulated trajectories as in
  Fig.~\ref{fig:Athens-xt}.}
\end{figure}

\subsection{\label{sec:Athens} Lane-based city traffic}

\begin{figure}[b]
\fig{0.95\textwidth}{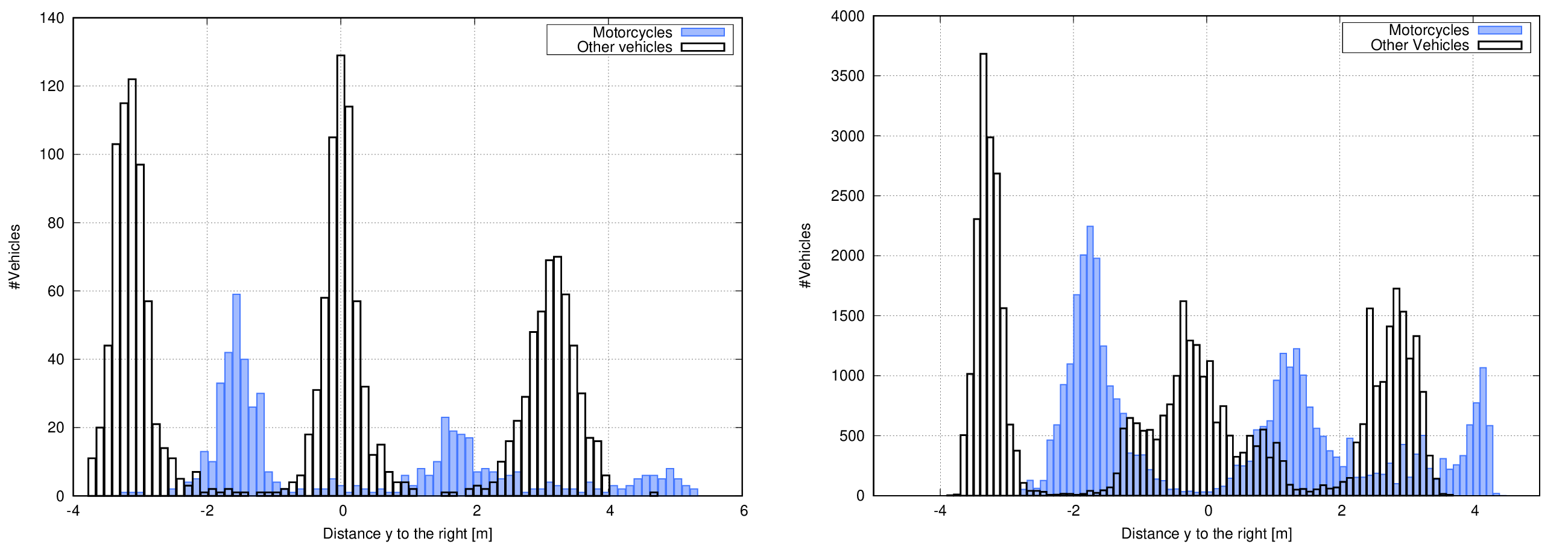}
\caption{\label{fig:Athens-hist}Histogram of the lateral positions of
  the motorcycles (blue) and the rest of the vehicles (black). Left:
real trajectories of the Athens arterial (see
Fig.~\ref{fig:Athens-xt}); right: simulation of the
  IAM.}
\end{figure}

The Figs.~\ref{fig:Athens-xt} (left), \ref{fig:Athens-xy} (left), and \ref{fig:Athens-hist}(left)
show naturalistic trajectory data of the
  second-to-right lane of the Athens arterial
  \emph{Leof. Alexandrias} as obtained by the open-data project
pNEUMA~\cite{pNEUMA} on Oct 24, 2018. The data show that the
motorcyclists generally make use of the space between the lanes
(Figs.~\ref{fig:Athens-xy} and ~\ref{fig:Athens-hist}) to overtake the
larger vehicles as seen in the $x$-$t$ cross section of the
trajectories at the middle lane and its boundaries
(Fig.~\ref{fig:Athens-xt}). This leads to a partial segregation of the
vehicle types with motorcycles accumulating behind the stopping lines
of the red traffic lights (red horizontal lines in
Fig.~\ref{fig:Athens-xt}) and starting first when the lights turn
green.

In the IAM (right plots of the respective figures), this particular
situation is 
modelled by floor fields with 
a phase shift of $\pi$ between motorcycles and other vehicles. The
simulation qualitatively and even semi-quantitatively reproduced all
observations such as the longitudinal and lateral vehicle-type
segregation, the longitudinal dynamics including the
overtaking maneuvers of the motorcyclists (Fig.~\ref{fig:Athens-xt}
right), the lane-changing rate (beginning and ending
trajectories in Fig.~\ref{fig:Athens-xt}, number of crossing trajectories in
Fig.~\ref{fig:Athens-xy}), and the lateral dynamics during lane
changes (Fig.~\ref{fig:Athens-xy}).

\subsection{\label{sec:bike} Lane-free bicycle traffic}

\begin{figure}[h]
\fig{0.95\textwidth}{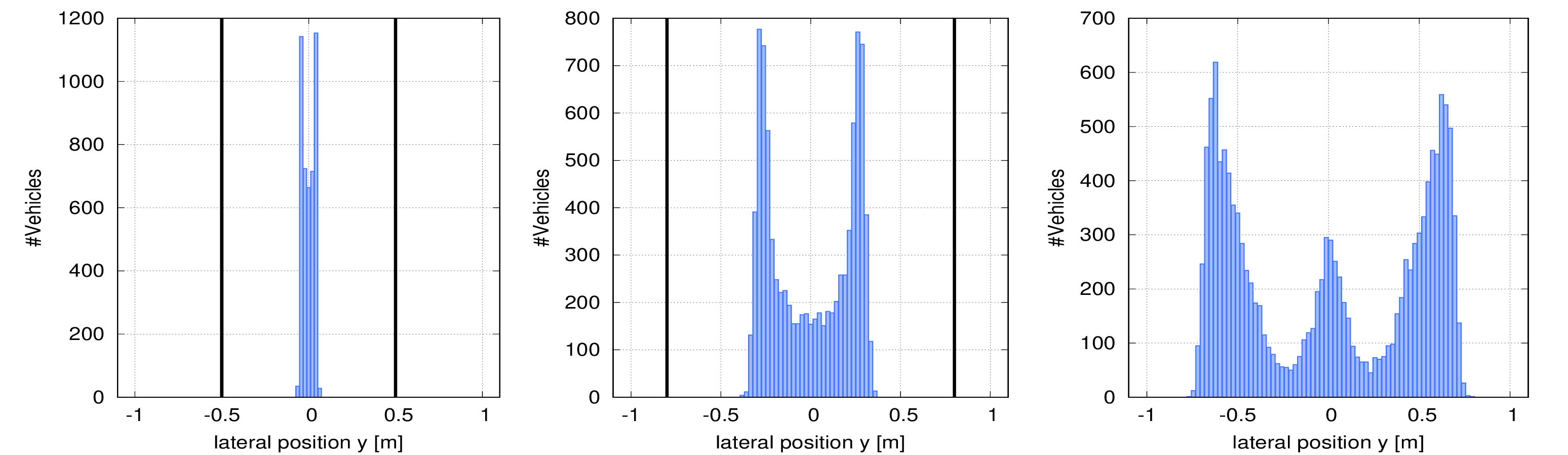}
\caption{\label{fig:bicycle-hist}Simulation of the lateral distributions of
  lane-free dense bicycle traffic
    for path width \unit[1.0]{m} (left), \unit[1.6]{m}(middle), and
    \unit[2.4]{m}(right).} 
\end{figure}

To demonstrate the ability of the IAM to simulate lane-free traffic
and various observed self-organisation
effects~\cite{wierbos2019capacity}, we simulate bicycle traffic on
bike paths of several widths. In order to create congested traffic, we
implemented a downstream bottleneck reducing the
capacity. For narrow paths (width
\unit[1]{m}), we observed
staggered single-lane following while two or free lanes emerge
for the  wider paths (Fig.~\ref{fig:bicycle-hist}). For even wider paths,
the lanes gradually vanish 
(not shown). For free traffic, the configuration is also different
with less emerging lines (two lanes, staggered or non-staggered single
lane) depending on the width and the traffic demand.
 
\begin{figure}[h]
\fig{0.95\textwidth}{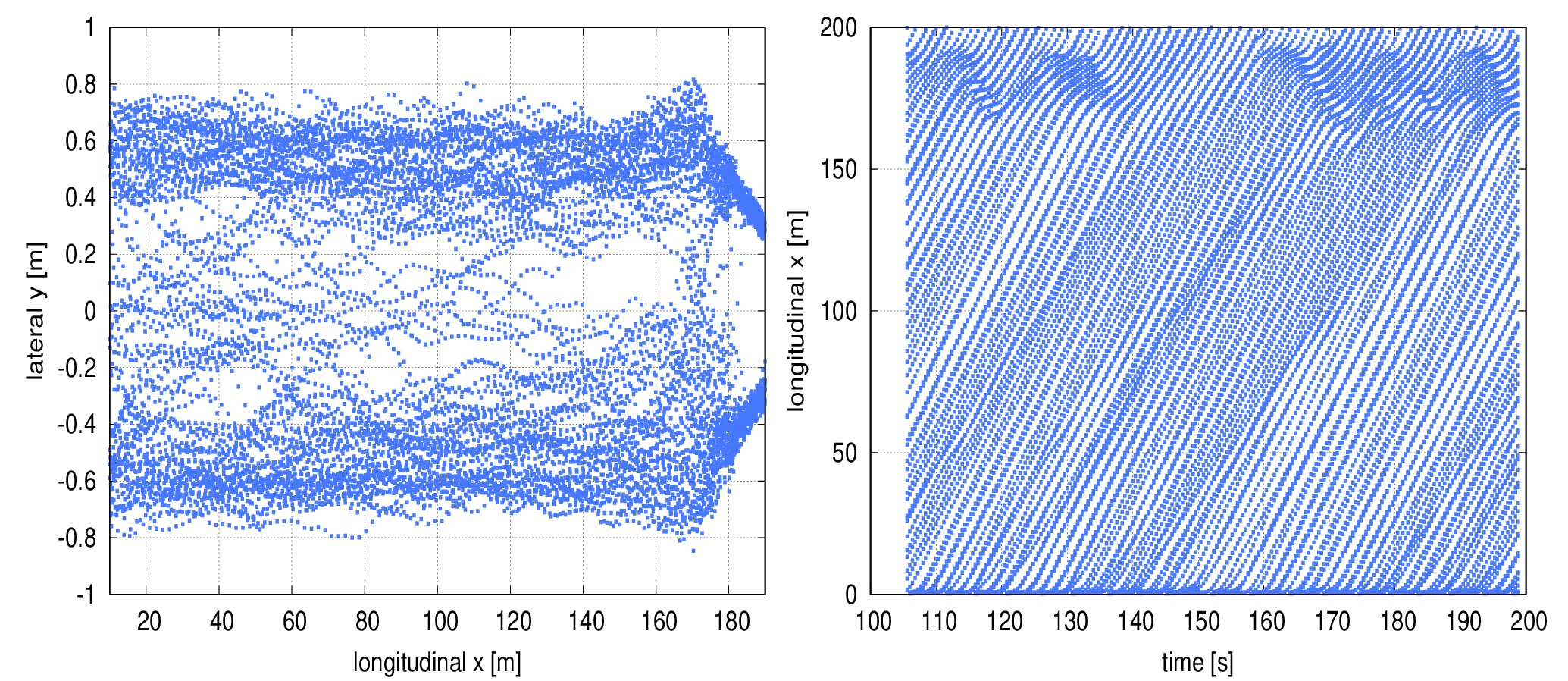}
\caption{\label{fig:bicycle-traj}Simulation of the trajectories of a
  \unit[2.4]{m} wide lane-less bike path in $x$-$y$ direction (left)
  and in $x$-$t$ direction (right).} 
\end{figure}

In Fig.~\ref{fig:bicycle-traj}, we take a closer look at the widest
considered bikepath ($w=\unit[2.4]{m}$) where a third center file
spontaneously begins to form. From $x=\unit[170]{m}$ to
$x=\unit[200]{m}$, a bottleneck in form of a gradual width reduction
to \unit[1.6]{m} is introduced and the three-lane traffic
spontaneously reorganizes into two-lane traffic, starting at
$x=\unit[140]{m}$, i.e.,
\unit[30]{m} \emph{upstream} of the beginning of the bottleneck.

\section{Discussion}
\label{sec:concl}
The proposed Intelligent-agent model (IAM) integrates the continuous
two-di\-men\-sio\-nal dynamics of the SFM with the high-speed properties of
car-following models to which it reverts for single-lane traffic. The
simulated 
anticipative high-speed behaviour of this model became most evident in the bicycle
simulation of Fig.~\ref{fig:bicycle-traj} where a spontaneous
transition of three-lane to two-lane flow occurs \unit[30]{m} upstream
of the bottleneck. This anticipation is driven by the relative-speed
term of the underlying IDM and can be observed for most longitudinal
sub-models containing a relative-speed term.  By construction, the SFM is not able to such an
anticipation.

 When
adding floor fields, we obtain an integrated car-following and
lane-changing model similar to
MOBIL~\cite{MOBIL-TRR07,traffic-simulation} but with continuous
lateral motion. We have shown that the IAM can describe the
observed dynamics and emergent phenomena of mixed vehicular and
motorcycle traffic and lane-free bicycle traffic. In the future, we
plan to validate the model on further trajectory data, including truly
lane-free mixed vehicular traffic~\cite{Kanagaraj2018self} 
and zipper merging and investigate
microscopic aspects of the cyclist's
configuration and how the capacity depends on the path
width~\cite{wierbos2019capacity}.  
For an online demonstration, see
\texttt{mtreiber.de/mixedTraffic/index.html}.

%

\bibliographystyle{elsart-num}





%

\bibliography{databaseLocal}

\end{document}